\documentclass[journal]{vgtc}                
\ifpdf
  \pdfoutput=1\relax                   
  \usepackage{graphicx}                
  \DeclareGraphicsExtensions{.pdf,.png,.jpg,.jpeg} 
\else
  \usepackage{graphicx}                
  \DeclareGraphicsExtensions{.eps}     
\fi%

\graphicspath{{figures/}{pictures/}{images/}{./}} 

\usepackage{microtype}                 
\PassOptionsToPackage{warn}{textcomp}  
\usepackage{textcomp}                  
\usepackage{times}                     
\usepackage{cite}                      
\usepackage{tabu}                      
\usepackage{booktabs}   

\usepackage{enumitem, kantlipsum}
\usepackage{caption}
\usepackage{subcaption}
\usepackage{xcolor}
\usepackage{colortbl}
\usepackage{float}

\usepackage{amsmath}
\usepackage{amssymb}
\usepackage[hidelinks]{hyperref}
\usepackage{soul}

\usepackage{adjustbox}
\newcolumntype{R}[2]{%
    >{\adjustbox{angle=#1,lap=\width-(#2)}\bgroup}%
    l%
    <{\egroup}%
}

\newcommand*\tablerot{\multicolumn{1}{R{90}{0em}}}
\definecolor{tablegray}{HTML}{F5F5F5}

\usepackage{color}

\definecolor{lhorange}{HTML}{bf5700}

\definecolor{table-orange}{HTML}{fb8072}
\newcommand{\yes}[1] {
    \raisebox{-0.1em}{\includegraphics[width=.9em]{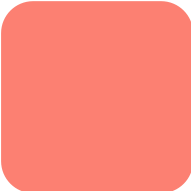}
}&}
\newcommand{\no}[1] {
    \raisebox{-0.1em}{\includegraphics[width=.9em]{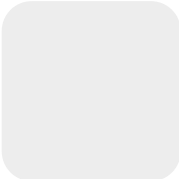}
}&}

\newif\ifnotes
\notestrue

\newcommand{\etc}[0]{etc.}
\newcommand{\eg}[0]{e.g.,}
\newcommand{\ie}[0]{i.e.,}
\newcommand{\knn}[0]{\textbf{\textsc{$k$-nn}}}
\newcommand{\bnb}[0]{\textbf{\textsc{b-bn}}}
\newcommand{\af}[0]{\textbf{\textsc{af}}}
\newcommand{\hmm}[0]{\textbf{\textsc{hmm}}}
\newcommand{\cm}[0]{\textbf{\textsc{cm}}}
\newcommand{\ad}[0]{\textbf{\textsc{ad}}}
\newcommand{\ac}[0]{\textbf{\textsc{ac}}}
\newcommand{\ens}[0]{\textbf{\textsc{ensemble}}}
\newcommand{\topk}[0]{top-$\kappa$}
\newcommand{\success}[0]{\textbf{\textsc{success}}}
\newcommand{\rank}[0]{\textbf{\textsc{rank}}}

\newcommand{\nexticon}[0]{\includegraphics[width=.018 \textwidth]{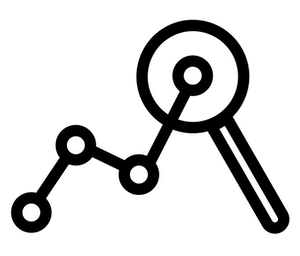}}
\newcommand{\biasicon}[0]{\includegraphics[width=.018 \textwidth]{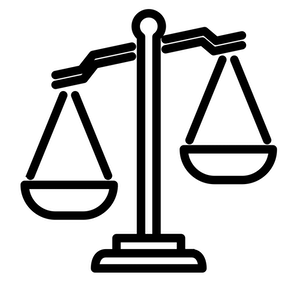}}
\newcommand{\histicon}[0]{\includegraphics[width=.018 \textwidth]{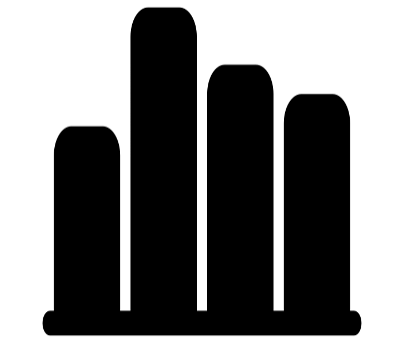}}
\newcommand{\disticon}[0]{\includegraphics[width=.018 \textwidth]{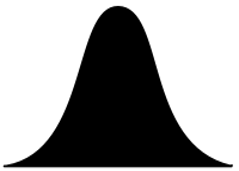}}
\newcommand{\texticon}[0]{\includegraphics[width=.0105 \textwidth]{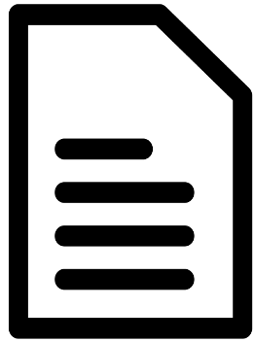}}
\newcommand{\ordinalicon}[0]{\includegraphics[width=.016 \textwidth]{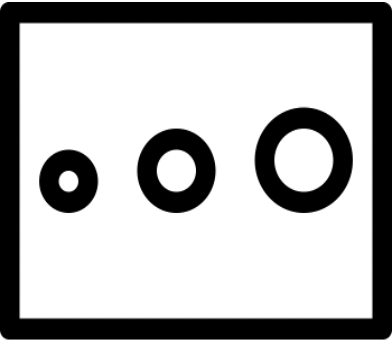}}

\onlineid{1618}

\vgtccategory{Research}
\vgtcpapertype{please specify}

\title{A Unified Comparison of User Modeling Techniques for \\ Predicting Data Interaction and Detecting Exploration Bias}

\author{Sunwoo Ha, Shayan Monadjemi, Roman Garnett, and Alvitta Ottley}
\authorfooter{
\item
 Sunwoo Ha is with Washington University. E-mail: sha@wustl.edu.
\item
 Shayan Monadjemi is with Washington University. E-mail: monadjemi@wustl.edu.
\item
 Roman Garnett is with Washington University. E-mail: garnett@wustl.edu.
 \item
 Alvitta Ottley is with Washington Universit. E-mail: alvitta@wustl.edu.
}

\shortauthortitle{Ha \MakeLowercase{\textit{et al.}}: A Unified Comparison of User Modeling Techniques}

\abstract{
The visual analytics community has proposed several user modeling algorithms to capture and analyze users' interaction behavior in order to assist users in data exploration and insight generation. For example, some can detect exploration biases while others can predict data points that the user will interact with before that interaction occurs. Researchers believe this collection of algorithms can help create more intelligent visual analytics tools. However, the community lacks a rigorous evaluation and comparison of these existing techniques. As a result, there is limited guidance on which method to use and when. Our paper seeks to fill in this missing gap by comparing and ranking eight user modeling algorithms based on their performance on a diverse set of four user study datasets. We analyze exploration bias detection, data interaction prediction, and algorithmic complexity, among other measures. Based on our findings, we highlight open challenges and new directions for analyzing user interactions and visualization provenance.
} 


\CCScatlist{ 
 \CCScat{K.6.1}{Management of Computing and Information Systems}%
{Project and People Management}{Life Cycle};
 \CCScat{K.7.m}{The Computing Profession}{Miscellaneous}{Ethics}
}

\begin{document}



\maketitle

\section{Introduction} 
\label{sec:intro}

Researchers in the visualization community have long viewed interaction as an \textit{analytic discourse} between the analyst and the visualization system~\cite{pike2009science}. Thus, capturing and analyzing the user's passive interactions has been integral to the visual analytics research agenda~\cite{cook2005illuminating,perry2009supporting,pirolli2005sensemaking,ribarsky2009science,crouser2013balancing}. Some believe that this data can provide a \textit{transcript} of the reasoning process, informing more effective visualization encodings, and producing better 
intelligent algorithms to assist with data exploration, model refinement, etc.~\cite{ragan2015characterizing, xu2020survey}. Further, many have 
seized the opportunity to leverage machine learning techniques to decode the information embedded in the user's interaction log data. This paper adopts the term used by Xu et al.'s~\cite{xu2020survey} recent survey on the analysis of user interactions and refers to the general goal of understanding the user and their sensemaking process as \textit{user modeling}.


Although there has been significant progress in developing algorithms that can reveal valuable information about the user and their analytic process, a unified comparison of the proposed techniques on different datasets, tasks, and analysis scenarios is lacking. Some suggested methods are presented theoretically without in-depth empirical evaluation, while others are validated using controlled user studies, though often with a single dataset. Furthermore, the algorithms are sometimes proprietary, and the community lacks benchmark datasets for easy comparison. These issues point to fundamental problems that can hinder research progress and question the practicality of user modeling techniques outside of academia.

To address this, we present a computational benchmark study, comparing previously proposed user modeling techniques from the visual analytics community. We first narrow the scope to real-time algorithms that learn from low-level interactions with data points that researchers can potentially use for providing real-time support during data exploration. Such user modeling techniques typically fall under two broad categories: (1) \textit{data interaction prediction}, \ie\ inferring data points that the analyst is likely to interact with in the near future and (2) \textit{exploration bias detection}, \ie\ detecting data features that are disproportionately explored by the user. These two categories, however, are not mutually exclusive as the successful detection of exploration bias might could also infer potential data points the user may interact with and vice versa. After surveying the body of work, we selected seven proposed techniques and standardized their input and output specifications to account for a variety of datasets. In addition to the selected models, we developed an ensemble model to predict data interaction and detect exploration bias. We then compared all eight modeling techniques' performance with four different publicly available user study datasets that vary in the number of visual attributes shown and the types of tasks the user study participants performed. 

We evaluate the user modeling algorithms based on the success rate of their data interaction prediction, their exploration bias detection, and computational complexity. In addition to the comparison, we highlight the open challenges we discovered. In evaluating these algorithms, we aim to provide a better understanding to researchers and practitioners interested in integrating existing user modeling techniques into their system and to encourage further advances in analyzing user interactions and visualization provenance. The implementations, pre-procesed data files, and links to the manuscripts and datasets used in this study will be freely available on GitHub \footnote{\url{https://github.com/washuvis/vis2022usermodels}} to support reproducibility. 

We summarize our contributions as follows:
\begin{itemize}[noitemsep]
    \item Using a unified notation, we provide an overview of existing modeling techniques in the visual analytics community for predicting data interaction and detecting exploration bias. 
    \item We compare and evaluate the performance of eight techniques on two aspects of user modeling with four unique user study interaction log datasets.
    \item Based on our evaluation, we provide recommendations and new research directions for analyzing user interactions and visualization provenance.
\end{itemize}

\section{Related Work}
\label{sec:related}


User modeling techniques can fall into broad categories based on shared objective and can exhibit extreme diversity, as presented by the extensive survey from Xu et al.~\cite{xu2020survey}. For example, researchers have proposed techniques to infer data interactions~\cite{monadjemi2020competing, ottley2019follow, healey2012interest}, various forms of bias~\cite{monadjemi2020competing, wall2017warning, gotz2016adaptive}, and even user attributes~\cite{brown2014finding,liu2020survey}. Additionally, other techniques aim to assist the user by data prefetching~\cite{battle2016dynamic, khan2019flux}, recommending visualizations~\cite{gotz2009behavior}, providing interface guidance~\cite{ceneda2016characterizing,ceneda2019review}, and improving data selection~\cite{fan2018fast,gadhave2021predicting}.
In our work, we aim to evaluate existing models within the categories of data interaction prediction and exploration bias detection. We selected these two categories because of their broad applicability to varying datasets and system designs.

\subsection{Exploration Bias Detection}
While exploring datasets and making decisions, humans are susceptible to cognitive limitations and biases that arise naturally from perception and intuition\cite{dimara2018task,ellis2018cognitive}. There are many different types of biases. For example, one form of bias, \emph{confirmation bias}, was described by Nickerson et al.~\cite{nickerson98confirmationbias} as interpreting evidence in ways that are partial to existing beliefs already in mind. Gigerenzer et al.~\cite{gigerenzer2009homo} defined \emph{cognitive bias} as a deviation between human judgment and a rational norm. Lee et al.~\cite{lee2019avoiding} define \textit{drill-down fallacies} as wrongfully attributing a deviation in trend to a local change, while it is in fact a more general phenomena. Cho et al.~\cite{cho2017anchoring} investigate the impact of \emph{anchoring bias} (i.e. focusing too much on one piece of information) while making decisions. 

This paper focuses on models for \textit{exploration bias}. Exploration bias is akin to \textit{sample selection bias}, a generic phenomenon in social sciences when the investigator fails to capture a random sample of the target population~\cite{winship1992models}. Much like sample selection bias can lead to biased inferences about a study population, exploration bias can result in false deductions from the data. Using this as motivation, Wall et al.~\cite{wall2017warning} introduced \textit{Attribute Distribution}, which is a metric for quantifying exploration bias by observing user interactions while inspecting individual data points. Their work aimed to draw the connection between cognitive biases and exploration bias (\ie\ cognitive biases can cause exploration bias) , with proposed subsequent solutions for increasing bias awareness during data analysis~\cite{narechania2021lumos}.

While exploration bias in the context of visual analytics often has a negative connotation, it may at times be intentional and be interpreted as a user's data interest \cite{wall2017warning}.
Other works have investigated several techniques to detect and quantify exploration bias a user may manifest while interacting with datasets.
For example, Gotz et al.~\cite{gotz2016adaptive} presented \textit{Adaptive Contextualization}, a statistical approach to detecting exploration bias by observing user interactions while filtering high-dimensional data.  
In their recent work, Monadjemi et al.~\cite{monadjemi2020competing} detected exploration bias via Bayesian model selection. 
While the motivation behind each of these modeling techniques may be different, they all share a critical characteristic: in essence, they all compare the distribution of the underlying data set with the distribution of data points with which the user has interacted.



\subsection{Data Interaction Prediction}

This paper also examines algorithms that predict users' data interactions. There has been several techniques for inferring which data points the user will likely explore in the future.
For example, Healy et al.~\cite{healey2012interest} dynamically identified and tagged data elements in their visualization that were of potential interest based on the user's actions using Bayesian classification. Battle et al.~\cite{battle2016dynamic} modeled a user's sequence of interactions to pre-fetch data for a map visualization. They maintained a Markov chain model of interactions, and ranked the data for pre-fetching according to the likelihood of the users taking actions corresponding to said data. Similarly, Ottley et al.~\cite{ottley2019follow} observed users' mouse interaction and succeeded in modeling the attention of users during visual data exploration with a hidden Markov model. Their algorithm leverages a user's observed clicks with a visualization system to obtain next click predictions. In addition to bias detection, Monadjemi et al.~\cite{monadjemi2020competing} showed that their framework can predict future interactions and summarize analytic sessions. Zhou et. al~\cite{zhou2021modeling} observed users to develop a model of their analytical focus and use that model to surface relevant medical publications to users during visual analysis of a large corpus of medical records. As evident in this brief overview, there are many variations of techniques that learns data points that may to relevant to a user by observing their data exploration. However, the visual analytics literature lacks a unified comparison among these techniques. 

\subsection{Evaluation of User Modeling Techniques}
There are many ways to evaluate an algorithm or technique, including complexity analysis, implementation analysis, and laboratory user study~\cite{munzner2009nested}. The choice of evaluation approach depends on the project goals and contributions. As a result, we observed a diverse set of validation methods ranging from crowdsourced user studies and case studies to interaction simulations (\eg\ \cite{ottley2019follow, zhou2021modeling, monadjemi2020active}). For this paper, we utilize user study datasets to assess how the current modeling approaches might perform under actual user data to gain insight into their potential use in a real-time system. Therefore, we review the prior work on evaluated user modeling techniques with user study datasets.



For example, Ottley et al.\cite{ottley2019follow} validated their model's ability to predict future interactions on one user study. While their study validated the technique successfully, it did not evaluate their technique against any baselines.
Zhou et al.\cite{zhou2021modeling} validated their technique's ability to identify relevant medical concepts via a user study.
They successfully validated their technique by comparing it to the ground truth elicited from the study participants and measured a set of usability metrics via a post-study survey. However, their analysis also did not include any baselines.
Additionally, Gotz et al.\cite{gotz2016adaptive} conducted a formal user study to investigate how their technique (Adaptive Contextualization) can mitigate selection bias. Their user study included a two experimental groups: one with access to bias mitigation features, and one without access to such features (baseline). With their work being one of the earlier work in detecting selection bias, they would not have been able to compare against other techniques that were proposed in later years.

Most similar to our work, Monadjemi et al.~\cite{monadjemi2020competing} evaluated the performance of their technique (Competing Models) with two existing techniques by Ottley et al.~\cite{ottley2019follow} and Wall et al.~\cite{wall2017warning} for data interaction prediction and bias detection, respectively. Their evaluation, while including baselines and multiple user studies, was still limited in that it did not include all existing techniques as baselines and did not consider as many user studies.
We extend upon their work by standardizing and comparing the performance of four additional user modeling techniques with two additional datasets. Moreover, we introduce an ensemble approach that combines the selected modeling techniques for both exploration bias detection and data interaction prediction. We hope that our analysis of these techniques will shed some light for researchers and practitioners who are interested utilizing them.

\begin{table*}[ht]
\centering
\setlength{\tabcolsep}{1pt} 
\renewcommand{\arraystretch}{1}
\footnotesize

    \caption{A summary of the characteristics of the user models in this study and their overall success on the tested datasets for next interaction prediction. \vspace{-1em}}
    \label{tab:methods-overview}
    \begin{tabular}{lp{0.02\textwidth}ccp{0.02\textwidth}ccp{0.02\textwidth}cccp{0.02\textwidth}ccp{0.02\textwidth}p{0.1\textwidth}p{0.1\textwidth}}

       &
       & 
       \multicolumn{2}{c}{ \cellcolor{table-orange} \textcolor{white}{\textbf{Goal}}}
       & &
       \multicolumn{2}{c}{ \cellcolor{table-orange} \textcolor{white}{\textbf{Data}}}
       & &
       \multicolumn{3}{c}{ \cellcolor{table-orange} \textcolor{white}{\textbf{Schema}}}
       & &
       \multicolumn{2}{c}{ \cellcolor{table-orange} \textcolor{white}{\textbf{Complexity}}}
       & &
       \multicolumn{2}{c}{ \cellcolor{table-orange} \textcolor{white}{\textbf{Overall Success}}}
       \\
       \\

        \textit{User Modeling Methods} 
        & 
        & \tablerot{\slshape Data Interaction}
        & \tablerot{\slshape Exploration Bias}
        & 
        & \tablerot{\slshape Continuous} 
        & \tablerot{\slshape Discrete} 
        & 
        & \tablerot{\slshape Preprocessing} 
        & \tablerot{\slshape Priors Required} 
        & \tablerot{\slshape Off the Shelf} 
        & 
        & {\slshape Inference}
        & {\slshape Prediction}
        & 
        & \textit{\slshape Accuracy} \\

        \hline 
        
        \ens 
        & 
        & \yes 
        & \yes 
        & 
        & \yes 
        & \yes 
        & 
        & \yes 
        & \yes 
        & \no 
        & 
        & $O(2^{d}t + 2^{d}d^{3})$ 
        & $O(nd^{2})$ 
        & 
        & 
        {\vrule}
        \raisebox{-.2em}{\includegraphics[height=1em, width=.082\textwidth]{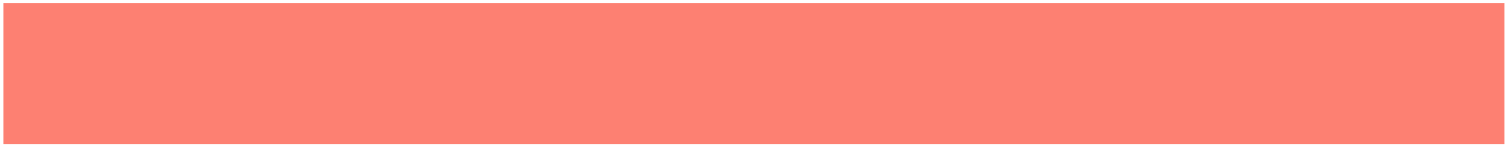}}
        \hspace{-.04\textwidth}  \textcolor{white}{\small \textbf{82}\%}
        \\

        \knn: $k$-Nearest Neighbors   \cite{monadjemi2020active} 
        & 
        & \yes 
        & \no 
        & 
        & \yes 
        & \yes 
        & 
        & \yes 
        & \no 
        & \yes 
        & 
        & \textit{None} 
        & $O(nd^{2})$ 
        & 
        & 
        {\vrule}
        \raisebox{-.2em}{\includegraphics[height=1em, width=.065\textwidth]{figures/bar.png} }
        \hspace{-.04\textwidth}  \textcolor{white}{\small \textbf{65}\%}
        \\

        \bnb: Boosted Naive Bayes \cite{healey2012interest} 
        & 
        & \yes 
        & \no 
        & 
        & \no 
        & \yes 
        & 
        & \yes 
        & \no 
        & \yes 
        & 
        & $O(nd)$ 
        & $O(nd)$ 
        & 
        & 
        {\vrule}
        \raisebox{-.2em}{\includegraphics[height=1em, width=.059\textwidth]{figures/bar.png}} 
        \hspace{-.04\textwidth}  \textcolor{white}{\small \textbf{59}\%}
        \\

        \af: Analytic Focus \cite{zhou2021modeling} 
        & 
        & \yes 
        & \no 
        & 
        & \no 
        & \yes 
        & 
        & \yes 
        & \no 
        & \yes 
        & 
        & $O(cd)$ 
        & $O(nd)$ 
        & 
        & 
        {\vrule}
        \raisebox{-.2em}{\includegraphics[height=1em, width=.059\textwidth]{figures/bar.png}}
        \hspace{-.04\textwidth}  \textcolor{white}{\small \textbf{59}\%}
        \\

        \hmm: Hidden Markov Model \cite{ottley2019follow} 
        & 
        & \yes 
        & \yes 
        & 
        & \yes 
        & \yes 
        & 
        & \no 
        & \yes 
        & \no 
        & 
        & $O(p)$ 
        & $O(np)$ 
        & 
        & 
        {\vrule}
        \raisebox{-.2em}{\includegraphics[height=1em, width=.067\textwidth]{figures/bar.png}}
        \hspace{-.04\textwidth}  \textcolor{white}{\small \textbf{67}\%}
        \\

        \cm: Competing Models \cite{monadjemi2020competing} 
        & 
        & \yes 
        & \yes 
        & 
        & \yes 
        & \yes 
        & 
        & \no 
        & \yes 
        & \no 
        & 
        & $O(2^{d}t + 2^{d}d^{3})$ 
        & $O(nd^{2})$ 
        & 
        & 
        {\vrule}
        \raisebox{-.2em}{\includegraphics[height=1em, width=.068\textwidth]{figures/bar.png}} 
        \hspace{-.04\textwidth}  \textcolor{white}{\small \textbf{68}\%}
        \\       
       
        \ad: Attribute Distribution \cite{wall2017warning} 
        & 
        & \no 
        & \yes 
        & 
        & \yes 
        & \yes 
        & 
        & \no 
        & \no 
        & \yes 
        & 
        & \textit{None} 
        & $O(d)$ 
        & 
        & 
        {\vrule}
        \\

        \ac: Adaptive Contextualization \cite{gotz2016adaptive} 
        & 
        & \no 
        & \yes 
        & 
        & \yes 
        & \yes 
        & 
        & \no 
        & \no 
        & \yes 
        & 
        & \textit{None} 
        & $O(d)$ 
        & 
        & 
        {\vrule}
        \\

    \end{tabular}
    \\ \vspace{1em}
    \fontsize{8}{9}\selectfont
    $c$ - \# of unique values for an attribute (specific to \af~\cite{zhou2021modeling}); $d$ - \# of data attributes;  $n$ - \# of points in the underlying dataset;\\ $p$ - \# of particles (specific to \hmm~\cite{ottley2019follow}); $t$ - \# of observed interactions;
\end{table*} 
\section{Purpose and Scope}
\label{sec:scope}

This paper presents a benchmark study to compare the user modeling methods proposed in the visual analytics literature, using publicly available user interaction datasets. As demonstrated in \autoref{sec:related}, the body of work on analyzing user interactions is diverse. 
However, instead of a comprehensive comparison of all existing user models, this paper focuses on techniques for predicting data interaction and detecting exploration bias. These categories of user modeling algorithms are complementary. For example, inferring the next data interaction could aid exploration bias detection, and we can use exploration bias inferences to identify data points relevant to the user's current tasks. Furthermore, we can observe from the literature that researchers have already developed algorithms that can already perform both tasks (\eg\ \cite{monadjemi2020competing,ottley2019follow}). 
Additionally, we selected these two user modeling categories because of their potential application in providing real-time support for data exploration. We pose the following research questions to guide our evaluations:
\begin{itemize}[noitemsep, topsep=0pt]
    \item []\textbf{RQ1:} Which algorithm is most accurate at predicting next data interactions for dataset with (i) goal-driven tasks and (ii) open-ended tasks? 
    \item []\textbf{RQ2:} Which algorithm is most accurate at detecting exploration bias for datasets with (i) goal-driven tasks and (ii) open-ended tasks?
\end{itemize}

\subsection{Algorithm and Dataset Selection}

To form our collection of algorithms and datasets discussed in this paper, we started with the pool of papers that were discussed in Xu et al.'s~\cite{xu2020survey} 2020 survey on the analysis of user interactions and provenance in visual analytics. The authors built a corpus of 105 manuscripts with seed papers collected from provenance-related prior work, supplemented by manually scanning all issues of major journals and all proceedings in the Visualization and HCI communities from 2009 to 2019. We accounted for the papers published around the same time and after the survey by performing a forward citation search on the manuscripts in the survey and the survey itself, resulting in a corpus of 110 manuscripts.

We narrowed the pool to the final selection of seven papers with algorithms within scope~\cite{ottley2019follow, wall2017warning, monadjemi2020competing, gotz2016adaptive, monadjemi2020active, healey2012interest, zhou2021modeling}.
To the best of our knowledge, this paper includes all real-time algorithms for detecting data interaction prediction and exploration bias that are (i) freely available, (ii) have precise code, or pseudocode descriptions, or (iii) can successfully be implemented following a reasonable amount of troubleshooting and debugging. A few edge cases could theoretically be included but are excluded from this analysis. For example, the \textit{Forecache} system proposed by Battle et al.~\cite{battle2016dynamic} captures movement data as a user navigates a large image, predicts where they will likely explore next, and uses this prediction to drive a prefetching algorithm. We can theoretically reformulate this problem and use the expected next move to infer the user's interaction, assuming the tasks is navigation-related. However, we did not include this or any other algorithm that required such reformulation. 

We used a similar process for our dataset selection. We manually examined our corpus of 110 papers to identify manuscripts with user studies and publicly available datasets. Four user study datasets logged participants' interactions with data elements throughout the study session. We included all four datasets in our evaluations.

\begin{table}[!b]
    \centering
    \caption{Notations used in this section.}
    \vspace{-.5em}
    \begin{tabular}{p{0.4\linewidth} p{0.51\linewidth}}
        \toprule
         \textit{Set Notation} & \textit{Description} \\
         \midrule
         ${\mathcal{D} = \{x_1, x_2, ..., x_n\}}$ & The set of data points visualized. \\
         \rowcolor{tablegray}
         ${\mathcal{A} = \{a_1, a_2, ..., a_d\}}$ & The set of data attributes. \\
         $\mathcal{E} = \{e_1, e_2, ..., e_m\}$   & The set of interaction types \newline (e.g.\ hover, click, etc.). \\
         \rowcolor{tablegray}
         ${\mathcal{I} = \{(x_{i, t}, c) \mid t = 1, 2, ...\}}$ & The sequence of interactions in a session. $(x_{i, t}, c)$ denotes taking action $c$ on data point $x_i$ at time step $t$. \\
         $f_{r} \colon \mathcal{D} \mapsto [0, 1]$ & The data ranking function for next data interaction prediction. $\mathcal{D}$, is mapped to a numerical value between 0 (low) and 1 (high). \\
         \rowcolor{tablegray}
         $f_{b} \colon \mathcal{A} \mapsto [0, 1]$ & The \emph{bias function} where the set of attributes, $\mathcal{A}$, is mapped to a numerical value between 0 (low) and 1 (high).\\
         \bottomrule
    \end{tabular}
    \label{tab:set_descriptions}
\end{table}

\section{User Modeling Methods}
\label{sec:methods}


In this section, we provide a brief overview on the existing modeling techniques considered in our benchmark study. For most of the selected modeling techniques, their implementations were not readily available. For ones that were readily available, the implementations were hard-coded to work with specific datasets. In order to easily evaluate the modeling techniques across multiple datasets, we implemented every technique with a standardized syntax and input/output format.

The \nexticon\ icon indicates that the technique is capable of data interaction prediction. The \biasicon\ icon indicates that the technique is capable of exploration bias detection. We compare a total of eight user models, based on a variety of theoretical approaches and summarized in \autoref{tab:methods-overview}. Of the algorithms we selected, three can only infer next data interaction (\textit{k}-Nearest Neighbors\cite{monadjemi2020active}, Boosted Naive Bayes\cite{healey2012interest}, and Analytic Focus \cite{gotz2016adaptive}), two can only detect exploration bias (Attribute Distribution~\cite{wall2017warning} and Adaptive Contextualization\cite{gotz2016adaptive}), and two can infer both (Hidden Markov Model\cite{ottley2019follow} and Competing Models\cite{monadjemi2020competing}). In addition to the selected algorithms, we also introduce and evaluate an ensemble model, developed by averaging the output of all modeling techniques.

Each model assumes there is a visualization of a dataset ${\mathcal{D} = \{x_1, x_2, ..., x_n\}}$, where each data point in $\mathcal{D}$ corresponds to a visual element. Each data point $x_i \in \mathcal{D}$ has $d$ attributes which may be continuous, categorical, or ordinal. Users may take different actions while interacting with data points using the visualization (i.e. click, hover, etc.). Thus, the models also assume a non-empty set of interactions supported by the system as $C = \{c_1, c_2, ...\}$. As users interact the visualized data points, they maintain a set of observation ${\mathcal{I} = \{(x_{i, t}, c) \mid t = 1, 2, ...}\}$, where $x_{i, t}$ denotes an interaction of type $c$ with data point $x_i$ at time step $t$.

\subsection{\textit{k}-Nearest Neighbors Classifier as seen in \cite{monadjemi2020active} \hfill \nexticon\ }

\noindent
A straightforward method for learning users' data interest by observing interactions is to train a $k$-nearest neighbors (\knn) binary classifier. 
\textbf{This approach assumes that proximity drives a user's exploration patterns.}
Monadjemi et al.\cite{monadjemi2020active} built a \knn classifier that relies on a notion of distance between data points and computing the matrix of $k$ nearest neighbors for every data point given the defined distance formula as a pre-processing step. The complexity for this pre-processing step is $O(n^{2}d)$. Refer to \autoref{tab:methods-overview} for more information and complexity variables. This model assumes that every data point in in the dataset, $x_i \in \mathcal{D}$, has a binary label $y_i \in \{0, 1\}$. 
A label of $y_i=1$ means the data point $x_i$ is relevant to the task at hand, whereas a label of $y_i=0$ means $x_i$ is irrelevant. By considering the labels of the nearest neighbors, this model is able to provide us with the probability of any given point being relevant in light of past observations: $\Pr(y_i=1 \mid x_i, \mathcal{I})$. Hence, we use this posterior belief as our ranking function:
\begin{equation}
    f_r(x_i) = \Pr(y_i=1 \mid x_i, \mathcal{I}).
\end{equation}


\subsection{Boosted Naive Bayes Classifier as seen in \cite{healey2012interest} \hfill \nexticon\ }

\noindent
Healey et al.\cite{healey2012interest} propose a naive Bayes classifier for maintaining a belief over users' interest in data points.
This approach tracks the frequencies of attributes explored and \textbf{assumes that the user's latent data interest is related to the dataset attributes' occurrence rate.}
In a similar formulation as the \knn\ approach, each data point $x_i \in \mathcal{D}$ has a binary label $y_i \in \{0, 1\}$. A label of $y_i = 1$ means the point $x_i$ is relevant to the task at hand and a label of $y_i = 0$ means it is not relevant. The goal of this approach is to reason about the unknown labels in light of frequencies of observations calculated via the Bayes' law:
\begin{equation}
    \Pr(y_i=1 \mid x_i) \propto \Pr(x_i \mid y_i = 1)\Pr(y_i=1).
\end{equation}
Extending $x_i$ to each of its individual $d$ dimensions and assuming conditional independence among attributes, we get:
\begin{equation}
\begin{split}
        &\Pr(y_i=1 \mid x_{i, 1}, x_{i, 2}, ..., x_{i, d})\\ 
        \propto &\Pr( x_{i, 1}, x_{i, 2}, ..., x_{i, d} \mid y_i = 1)\Pr(y_i=1) \\
        = & \Pr(y_i=1) \prod_{j=1}^{d} \Pr(x_{i,j} \mid y_i=1),
\end{split}
\end{equation}
where $x_{i, j}$ denotes the $j$th attribute of the $i$th data point. Therefore, we get the ranking function as:
\begin{equation}
    f_r(x_i) = \Pr(y_i=1) \prod_{j=1}^{d} \Pr(x_{i,j} \mid y_i=1).
\end{equation}

While Naive Bayes often performs as well as more sophisticated models in practice \cite{langley1992analysis}, its performance can further be improved by \emph{boosting}. As recommended by Healey et al.\cite{healey2012interest}, we used AdaBoost to train a stronger classifier.


\subsection{Analytic Focus Modeling as seen in \cite{zhou2021modeling} \hfill \nexticon\ }

\noindent
Zhou et al.\cite{zhou2021modeling} propose a technique for modeling \textit{analytic focus} (\af) during a session. Their technique is based on an abstract notion of \emph{concepts} defined as \emph{``meaningful data attribute[s] in the problem domain."} This technique tracks user focus on each of the concepts by observing their interactions and maintain an importance score for each concept.  \textbf{The model assumes that the user's latent data interest is related to the occurrence rate of discrete \textit{concepts} observed in their data interactions.}
With concepts being left open for practitioners to define in their particular domain, here, we let each unique value appearing in the discrete attributes of the dataset to represent a concept, performing a pre-processing step on continuous attributes to convert them into discrete bins. 

Each interaction $\alpha$ is defined by two values: an initial importance score $I_\alpha(0)$ and a persistence score $P_\alpha$. In a sense, $I_\alpha(0)$ refers to how intentional an action is and how informative it is in revealing analytic focus while $P_\alpha$ is a measure of how long the visual changes resulting from action $\alpha$ remain.
Using these parameters and the Ebbinghaus forgetting curve \cite{ebbinghaus1880urmanuskript,ebbinghaus1885gedachtnis,murre2015replication}, the authors define the per-action importance score function $t$ time steps after an action is taken as: 
\begin{equation}
    I_{\alpha}(t) = I_{\alpha}(0) \times e^{-\frac{t}{P_{\alpha}}}.
\end{equation}

Using the equation above, the authors propose an additive model for calculating the importance score for each concept $c$ at time $\tau$:
\begin{equation}
    I^c(\tau) = \sum_{(x_{i, t}, \alpha) \in \mathcal{I}_{\tau,c}}I_{\alpha} (\tau - t),
\end{equation}
where $\mathcal{I}_{\tau, c}$ denotes the set of interactions with points involving concept $c$ up until time step $\tau$. Considering that each data point $x_i \in \mathcal{D}$ may subscribe to multiple concepts (one concept per dimension in our case), we extend this method so that the importance of a data point is the product of the importance of its individual concepts. In other words, given a data point $x_i \in \mathcal{D}$ and the current time step $\tau$, 
\begin{equation}
    f_r(x_i) = \prod_{c \in C(x_i)}{I^c(\tau)},
\end{equation}
where $C(x_i)$ is the list of concepts with which $x_i$ is associated.



\subsection{Hidden Markov Model as seen in \cite{ottley2019follow} \hfill \nexticon\ \biasicon\  }

\noindent
Ottley et al.\cite{ottley2019follow} propose a \textit{Hidden Markov Model} (\hmm) approach for modeling user attention during visual exploratory analysis. \textbf{Their approach assumes that the visual features of the interface will drive the user's attention (it considers all possible combinations)} and their interactions such that attention at time $t+1$ will be similar to the observation at time $t$ . They constructed a hidden Markov model, presuming the user's behavior evolves under a Markov process (that is, the behavior at a particular time only depends on their behavior at the previous time step), and interaction events are generated conditionally independently given this sequence of attention shifts. The set of visible visual features (\eg\ color, position, size, etc.) creates the \textit{mark space} and the latent states. Further, their algorithm uses particle filtering~\cite{gordon1993novel,doucet2000sequential} to sample from the model's posterior distribution and predict the user's attention for the next time step. Integral to this technique is a \textit{bias vector} that captures the relative importance of the mark space components. We use the particle filter output to predict next data interaction and the bias vector for bias detection.

\begin{figure*}[!ht]
    \centering
    \includegraphics[width=0.95\linewidth]{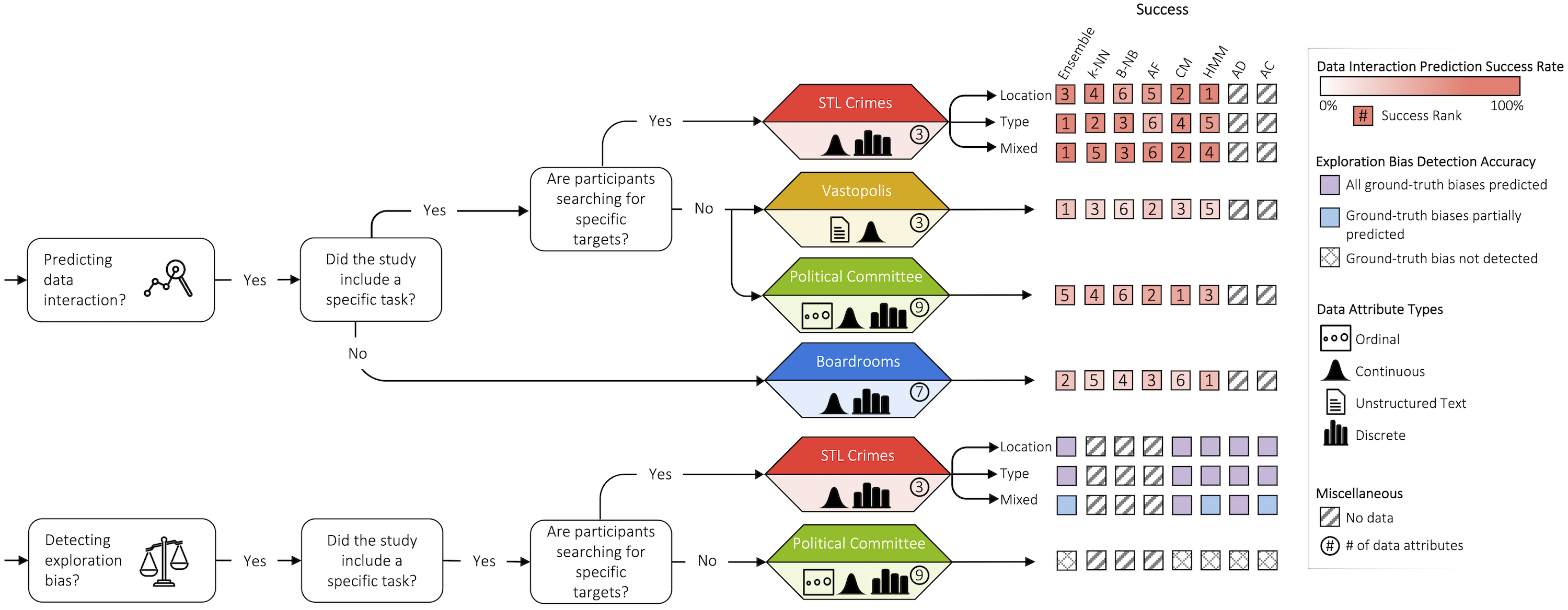}
    \caption{A summary of the characteristics of the selected user interaction datasets and the performance of the user modeling techniques with said datasets. We represent success using each technique's performance at top-100 points for next data interaction prediction.}
    \label{fig:datasets}
\end{figure*}


\subsection{Competing Models as seen in \cite{monadjemi2020competing} \hfill \nexticon\ \biasicon\ }

\noindent
Monadjemi et al.\cite{monadjemi2020competing} propose a Bayesian model selection approach for detecting exploration bias and inferring next data interaction during visual exploratory analysis. This approach, called \emph{Competing Models} (\cm), enumerates a set of probabilistic models each of which represent exploration based on a subset of data attributes. 
Given a dataset with $d$ attributes, they enumerate the model space ${\mathbf{M} = \{\mathcal{M}_1, \mathcal{M}_2, ..., \mathcal{M}_{2^d}\}}$ where each model represents exploration based on one subset of attributes. \textbf{Thus, it assumes and one or more data attributes (it considers all possible combinations) are sufficient to represent a users data interest.} As interactions arrive, they maintain a belief ${\Pr(\mathcal{M}_i \mid \mathcal{I})}$ which is interpreted as the viability of model $\mathcal{M}_i$ in explaining user interactions.
To rank the data points to the task at hand in light of past interactions, they use Bayesian model averaging to get:
\begin{equation}
    f_r(x_i) = \Pr(x_i \mid \mathcal{I}) = \sum_{j=0}^{2^d}\Pr(x_i \mid \mathcal{I}, \mathcal{M}_j).
\end{equation}
To quantify bias towards each attribute, they use the law of total probability as follows:
\begin{equation}
    f_b(a_i) = \sum_{\mathcal{M}_j \in \mathbf{M}_{a_i}} \Pr(\mathcal{M}_j \mid \mathcal{I}),
\end{equation}
where $\mathbf{M}_{a_i} \subset \mathbf{M}$ denotes the set of models involving attribute $a_i$.


\subsection{Attribute Distribution as seen in \cite{wall2017warning} \hfill \biasicon\ }

\noindent
Wall et al.\cite{wall2017warning} propose a set of metrics for quantitatively detecting various types of cognitive biases. The most relevant to our work is the metric they call \emph{attribute distribution}. 
\textbf{Their approach assumes that in an unbiased exploratory session, the distribution of values explored by the user will resemble the distribution of the underlying data.} It relies on well-known statistical tests, Chi-Square\cite{mchugh2013chi} and Kolmogorov–Smirnov\cite{massey1951kolmogorov}, which are used for discrete and continuous attributes respectively. In each case, the test takes the underlying data as well as the set of points with which the user has interacted as inputs and returns a $p$-value indicating whether or not the two samples could be reasonably drawn from the same distribution. Then, they define their attribute distribution metric for a given attribute $a_i \in \mathcal{A}$ to be:
\begin{equation}
    f_{b}(a_i) = 1 - p\text{-value}_{a_{i}}.
\end{equation}


\subsection{Adaptive Contextualization as seen in \cite{gotz2016adaptive} \hfill \biasicon\ }

\noindent
Gotz et al.\cite{gotz2016adaptive} propose \textit{Adaptive Contextualization}, a metric to measure selection bias in real-time. In order to detect such biases, they propose measuring the Hellinger distance between the datasets before and after filtering. The Hellinger distance is a statistical measure designed to quantify the similarity between two distributions. Thus, \textbf{this approach assumes that we can detect bias by comparing the observed distribution to the expected data distribution}. The authors propose using the Hellinger distance defined for discrete probability distributions with a pre-processing step to discretize the continuous attribute into discrete bins. 

Consider a particular dimension of interest, $a_i$, represented as a set of $m$ discrete values $\{v_1, v_2, ..., v_m\}$. The distribution of these values in the underlying data and interaction data are $\{p_1, p_2, ..., p_m\}$ and $\{q_1, q_2, ..., q_m\}$ respectively.  We compute the Hellinger distance as:
\begin{equation}
    f_b(a_i) = H(\mathcal{D}_{a_i}, \mathcal{I}_{a_i}) = \sqrt{\frac{1}{2} \sum_{j=1}^{m}{(\sqrt{p_j} + \sqrt{q_j})^2} },
\end{equation}
where $\mathcal{D}_{a_i}$ and $\mathcal{I}_{a_i}$ denote only the $a_i$ attribute of the underlying and interaction datasets respectively. While this technique was intended to detect selection bias (i.e.\ before vs.\ after filtering), we adopt it to quantify exploration bias.




\subsection{Ensemble Approach \hfill \nexticon\ \biasicon\ }


We anticipate that no model will be able to correctly recognize relevant points in every system, task, or dataset. This phenomena was also observed by the machine learning community which led to the development of \emph{ensemble methods} \cite{rokach2010ensemble,dietterich2000ensemble}. Ensembles are a collection of classification models combined to construct a model with higher predictive power. Using this approach, we consider a final technique which combines predictions from all five techniques that predict data interactions and all four techniques that detect exploration bias reviewed in this paper. To predict next data interaction, for a given data point, $x_i \in \mathcal{D}$, we compute the ranking function, $f_r(x_i)$, using each of the modeling techniques and average the ranking values to get the ensemble prediction. Then, we proceed with ranking the data points, a process identical to the one in the preceding modeling techniques. Similarly, we average the bias values towards each data attribute from each modeling technique to get the ensemble prediction for exploration bias.

\section{User Study Interaction Logs}
\label{sec:datasets}
 We used four publicly available interaction logs from user study experiments with visualization systems. These studies contained varying interfaces as well as task designs. We aimed to evaluate and compare the performance of the user modeling techniques with a variety of interaction logs, ranging from studies with very directed and specific tasks to the polar opposite, open data exploration. Refer to ~\autoref{fig:datasets} for a detailed summary of the characteristics of each dataset.


\subsection{\textbf{STL Crimes} as seen in \cite{ottley2019follow} \hfill \disticon\ \histicon\ }
The \textit{STL Crimes} dataset \textbf{involved a highly specified task thus capturing less exploration noise. We selected this dataset as it represents the best-case observational scenario for user modeling.}

Curated by Ottley et al.~\cite{ottley2019follow}, 
this dataset consists of 1,951 crime instances reported in the month of March 2017 with a total of eighteen attributes. Each instance of crime on the map was displayed as a dot with a position and color, indicating the location and type of the crime respectively. Eight types of crime exist in the database: homicide, theft-related, assault, arson, fraud, vandalism, weapons, and vagrancy.  30 participants were recruited from Mechanical Turk and they completed three categories of directed search tasks that were designed to encourage them to either click on: (1) \textit{Type-Based} data points with similar crime categories (e.g. Homicide, Assault, etc.), (2) \textit{Location-Based} data points that are within in the same vicinity, or (3) \textit{Mixed} data points that are of the same crime category and in the same vicinity. Each participant completed a total of six tasks, two from each category. Since the authors specified to the participants target attributes to interact with, this provides us with a notion of ground-truth bias. In their user study, Ottley et al.~\cite{ottley2019follow} captured mouse click data as participants interacted the visualization. Consistent with Ottley et al.~\cite{ottley2019follow}, we filtered the data to hold only the interactions of participants who successfully answered the task questions and created three data subsets: 28 user sessions for location-based tasks, 23 for typed-based, and 27 for mixed. 


\subsection{\textbf{Vastopolis} as seen in \cite{monadjemi2020active} \hfill \texticon\ \disticon\ }
We selected the \textit{Vastopolis} dataset because of the attributes in the dataset. \textbf{It is the only user interaction dataset in this study that includes unstructured text.}

Inspired by the 2011 VAST Challenge, Monadjemi et al.'s~\cite{monadjemi2020active} study describes a major epidemic that started in the fictitious city of Vastopolis. Participant's interacted with a map of Vastopolis with 3000 geolocated tweet-like data. In addition to the social media posts, the map display major roadways, waterways, and landmarks in the city. They recruited 130 participants from Amazon Mechanical Turk and tasked each participant to search through the dataset of microblogs via an interactive map and bookmark as many posts containing illness-related information as possible. Following the same pre-processing step as the authors, 74 user interaction logs remained from the control group of the study. The only interaction type we considered in the interaction logs were intentional ``bookmarks" of the microblogs.

\subsection{\textbf{Political Committee} as seen in \cite{wall2022left}\hfill \ordinalicon\ \disticon\ \histicon\ }
The \textit{Political Committee} user study \textbf{examined cognitive and exploration bias as participants selected their picks for a hypothetical committee. Thus we selected this dataset as a candidate for examining exploration bias.}

Wall et al.~\cite{wall2022left} generated a dataset of 180 fictitious politicians, representing the composition of the Georgia General Assembly. The dataset contains three discrete and six continuous attributes that characterizes each politician. In the study conducted by Wall et al.~\cite{wall2022left}, the authors had users interact with an visualization system that supports the exploration of the fictitious political committee data. There were two versions of the visualization system: a Control version of the interface, and an Intervention version of the interface, which was modified to visualize traces of the user’s interactions with the data in real-time. To assess the effectiveness of visualizing the user's interaction traces, the authors performed an in-lab study with 24 participants. They were tasked with selecting a committee of 10 candidates to review public opinion in Georgia on the controversial bill that bans abortion after 6 weeks. In order to observe the raw, inherent biases of the users, we selected the interaction logs of those in the Control group (total of 12 participants). The authors recorded several types of interactions the user can make with the interface such as hovers, changing of axes on the scatterplot, committee selection etc. We filtered the interaction log for each user to contain only committee selection interactions with the data points.


\subsection{\textbf{Boardrooms} as seen in \cite{feng2018effects}\hfill \disticon\ \histicon\ }
The \textit{Boardrooms} study did not involve a particular task, but instructed the participants to freely interact with the visualisation. \textbf{We expect this dataset to include the highest levels of exploration noise. }

Feng et al.~\cite{feng2018effects,feng2018patterns} utilized the visualization, ``Inside America’s Boardroom," which was published by the \textit{Wall Street Journal}~\cite{wsjboardrooms} in 2016. This point-based visualization displays seven different attributes to the user: market capitalization, ratio of unrelated board members, ratio of female board members, average age of board members, average tenure of board members, and median pay of board members. The study adopted an open-ended approach and observed the user's interactions as they explored the dataset through the visualization.  The interaction log contains a succession of hovers each user made on different data points. To account for unintentional hovers, we followed the same approach as Monadjemi et al.~\cite{monadjemi2020competing}, each user's session was filtered only include hovers that lasted for over one second. Additionally, we only considered sessions that consisted of more than three hovers, resulting in a total of 39 user sessions.

\section{Evaluation Measures}
In this section, we describe the performances measures we used to evaluate the modeling techniques' accuracy in predicting data interactions and detecting exploration bias with the four user study logs. 
We note that 2 out of 4 interaction logs included a single interaction type. In particular, the \textit{STL Crimes} captured only mouse clicks and \textit{Boardrooms} recorded mouse hovers alone. \textit{Vastopolis} included bookmarks and hovers, but recording inconsistencies in the logging made it impossible to integrate the two interaction modalities. Thus, our evaluation focused solely on predictions based on the more intentional ``bookmark'' interaction. Finally, the \textit{Political Committee} dataset included the most diverse interface and data interactions  ( \eg\ hovers, reconfiguring, committee selections). For consistency, we select a single interaction observation (committee selections), and the algorithms’ predictions are limited to their observations. Suppose, for example, the algorithm’s observations are limited to mouse clicks. Then the algorithm can only predict the next click or detect data exploration bias among the clicked data points. In the case of \textit{Political Committee}, this means that the algorithms can only observe when the user selects a committee member and must make predictions about the next committee selection or estimate selection biases solely from the limited observations.

\subsection{Evaluating Data Interaction Prediction}
We developed a standardized probabilistic output for each algorithm to compare the models to each other. Specifically, the techniques treat every data point as a potential candidate for being most relevant to the user's data interest. Thus, each method outputs the ranking probability for each data point given its observations. This output represents the technique's unique interpretation of the data points that the user will likely interact with. For example, \knn\ assumes that the data points most relevant to the user will minimize some measure of distance while \af\ examines the frequency on \textit{concepts} from the past interactions. 

Given the model's probabilities, we can rank each data point based on the algorithm's belief.
We report \topk\ data retrieval for the algorithms for $\kappa \in \{1,5,10,20,50,100\}$ which we will refer to as the \textit{prediction sets}. We considered a variety of set sizes. For example, an alternative is to use percentages instead of a fixed values. However, for large datasets, retrieving a small percentage to the dataset can still be overwhelming to the user if we consider the goal of provided real-time analysis support. This standardized output enables us to evaluate and compare each technique's ability to predict the next data interaction. We record the following measures for every interaction the datasets:
\begin{itemize}[noitemsep, topsep=0pt]
    \item \textbf{\success:} $\in \{0,1\}$, binary value for whether the next data interaction was included in the prediction set for all values of $\kappa$.
    \item \textbf{\rank:} the ranked position of the next observed interaction such that lower is better.
\end{itemize}



\begin{figure}[b!]
    \centering
    \includegraphics[width=0.9\linewidth]{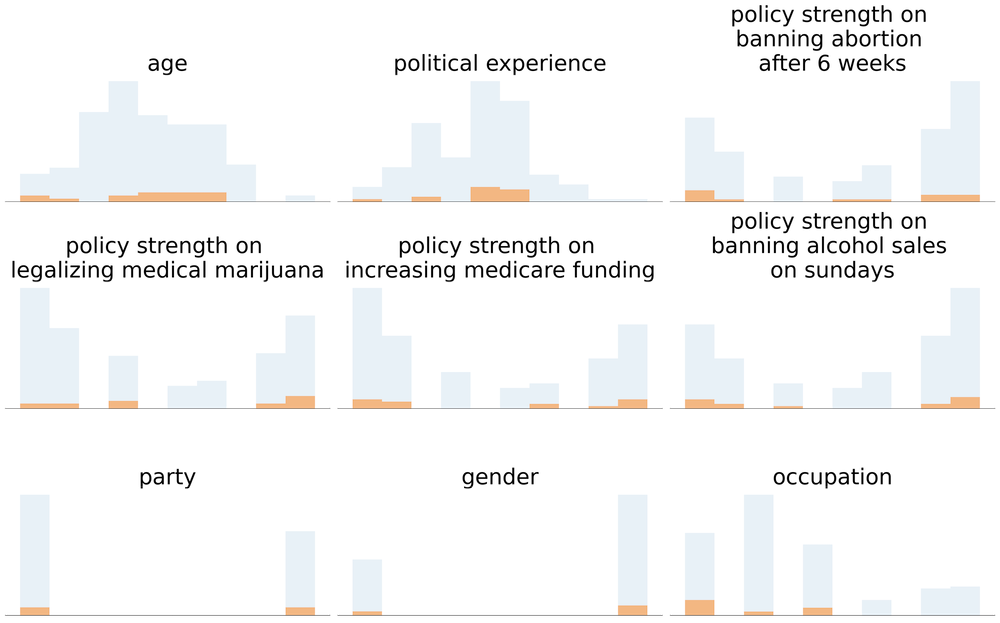}
    \caption{Comparing participant \textit{Lima's} interactions (foreground) to the full dataset distribution (background). Both coders reported no significant difference between the two distributions, indicating no bias detected. The result were similar for all 12 study participants from \cite{wall2022left}.}
    \label{fig:exampleuser}
\end{figure}

\begin{figure*}[!ht]
     \centering
     \begin{subfigure}[b]{0.87\textwidth}
         \centering
         \includegraphics[width=\textwidth]{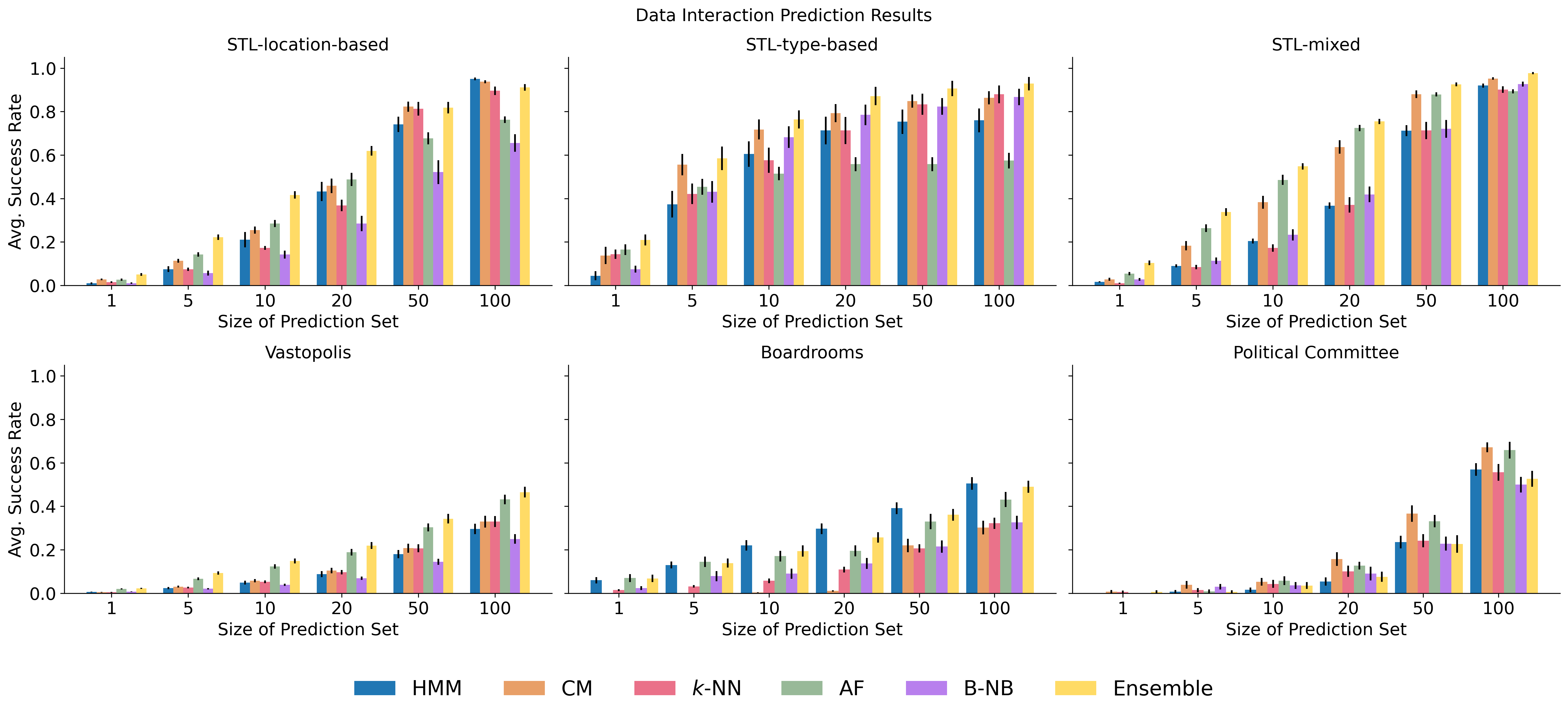}
         \vspace*{3mm}
     \end{subfigure}
     \begin{subfigure}[b]{0.87\textwidth}
        \vspace{-1em}
         \centering
         \includegraphics[width=\textwidth]{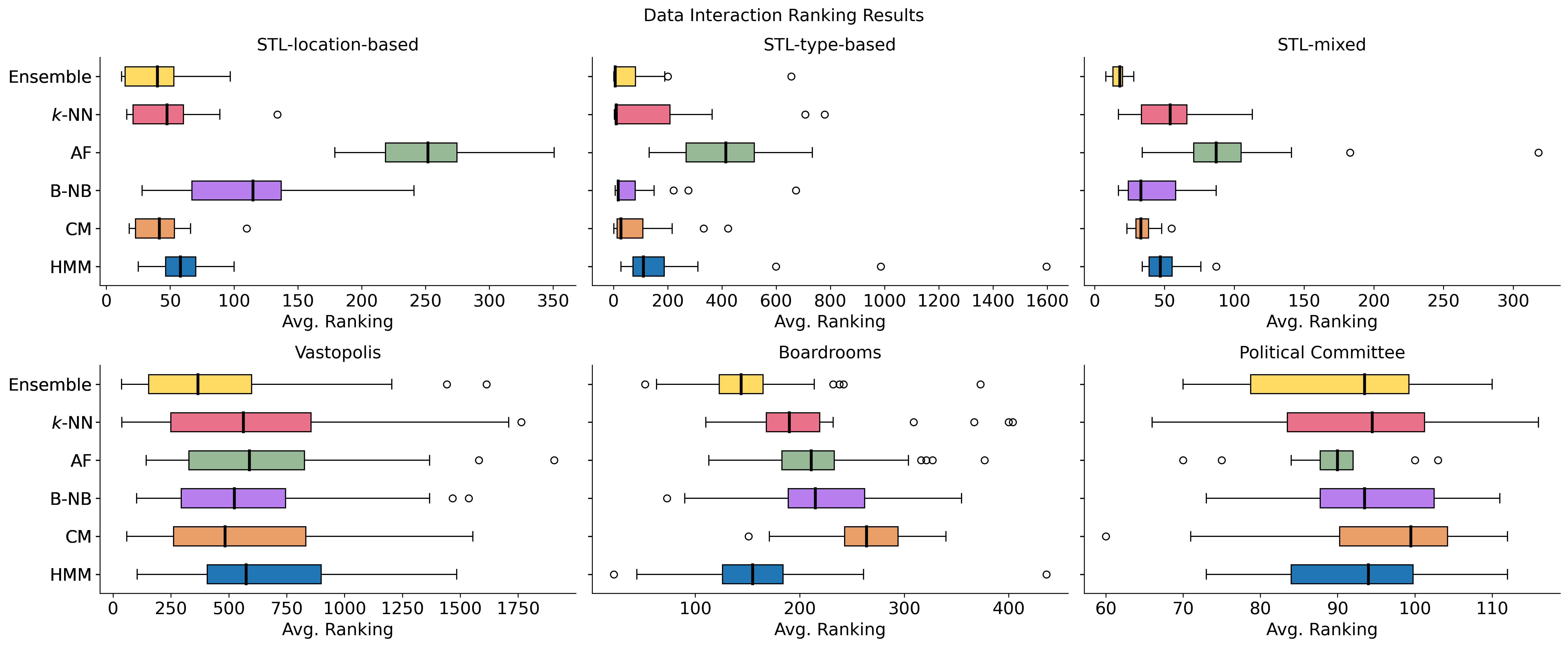}
     \end{subfigure}
     \caption{We evaluated each model's performance with two metrics: (top) average \textbf{success rate} across all the tasks in each user study dataset for varying values of the prediction set size and (bottom) average \textbf{rank} of the next data interaction of the user modeling techniques.}
     \label{fig:inference}
     \vspace{-2mm}
\end{figure*}

\subsection{Evaluating Exploration Bias Detection}
\label{sec:bias_detection_gt}

Evaluating each technique's ability to detect exploration bias poses a unique challenge, in that \textit{ground truth} is not always readily available. When evaluating data interaction prediction, we had access to a notion of \emph{success}: we wanted to predict next data points with which the user interacts, and we were able to verify our ability to do so by observing the data interaction point. For bias selection, however, such ground truth is not as easy to access, especially for user study datasets with open-ended tasks. With the \textit{Political Committee} dataset collected by Wall et al.~\cite{wall2022left}, for example, each user can exhibit their own set of biases as they explore the data. 

For this set of evaluations, we use two datasets: \textit{STL Crimes} separated into \textit{location-based}, \textit{type-based}, and \textit{mixed} subsets and \textit{Political Committee}. The \textit{STL Crimes} user study involved a directed task, asking participant to interaction with specific portions of the data, inherently resulting in biased data exploration. For example, the type-based task instructed participants to inspect all ``Arsons'' and note the time of day they occurred. We used the specific task as the ground truth. In our evaluation, an algorithm successfully detects exploration bias if its outputs indicate data attributes related to crime location, type, or a mixture model of type and location for \textit{location-based}, \textit{type-based}, and \textit{mixed} respectively. 

For \textit{Political Committee}, two researchers performed qualitative coding on each study participants' committee selections, following best practices~\cite{brod2009qualitative}. We used an iterative approach. First, the coders reviewed a single participant's selections and density plots showing the attribute distributions for a given set of interaction logs and the original data distribution. Their goal was to identify evidence of data bias in the observed selections. Evidence of data bias was determined if the attribute distributions for a given set of interaction logs was different than original data distributions. Next, the coders established a codebook from this first round and independently coded all twelve participants' interaction logs. After inspecting each of the twelve study participants, both coders agreed that there was no observable bias in the any of the interaction logs. For example, \autoref{fig:exampleuser} compares study participant Lima's data selection to the individual attribute distribution. No significant difference was observed between the two distributions for each attribute. Although the coding sessions detected no bias, we included this analysis to assess the algorithms' performance in detecting bias when none exists.


\begin{figure*}[!ht]
     \centering

     \begin{subfigure}[b]{\textwidth}
         \centering
         \includegraphics[width=0.87\textwidth]{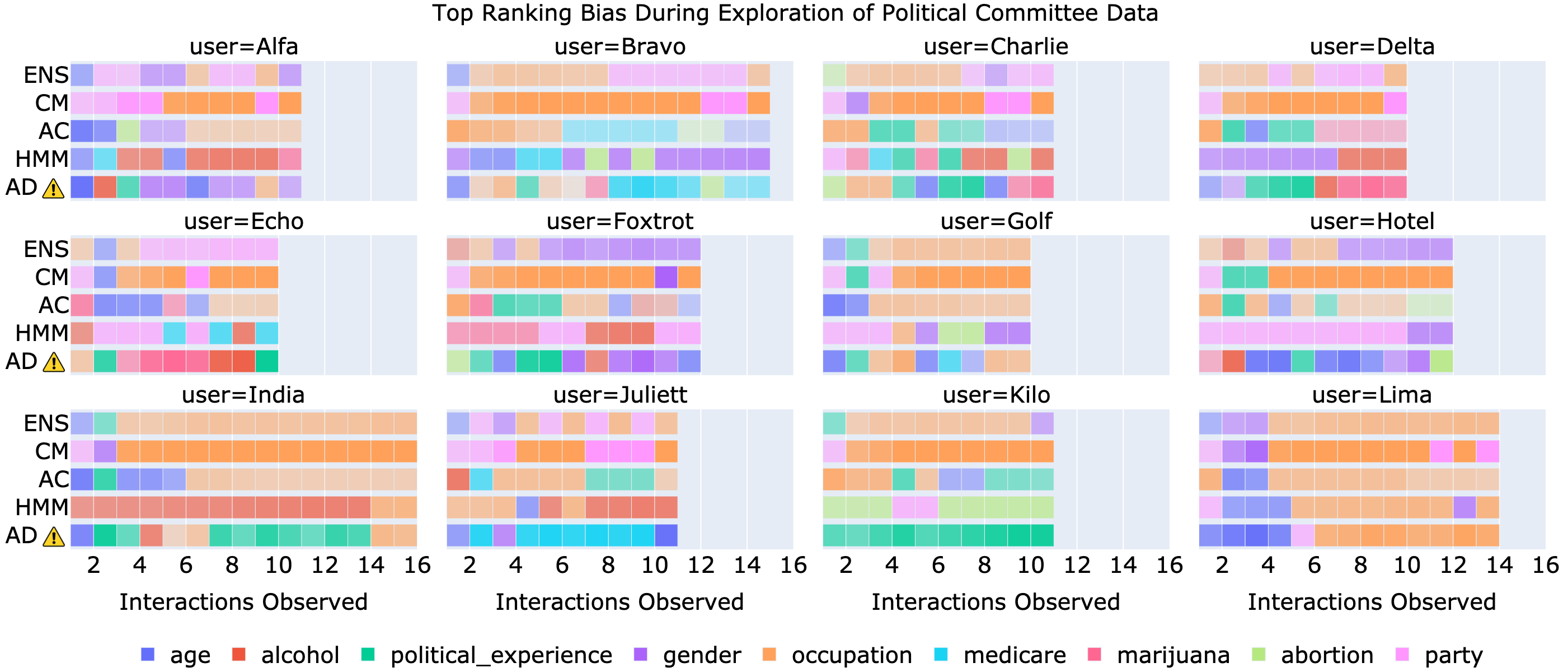}
     \end{subfigure}
     \begin{subfigure}[b]{\textwidth}
        \centering
        \includegraphics[width=0.87\textwidth]{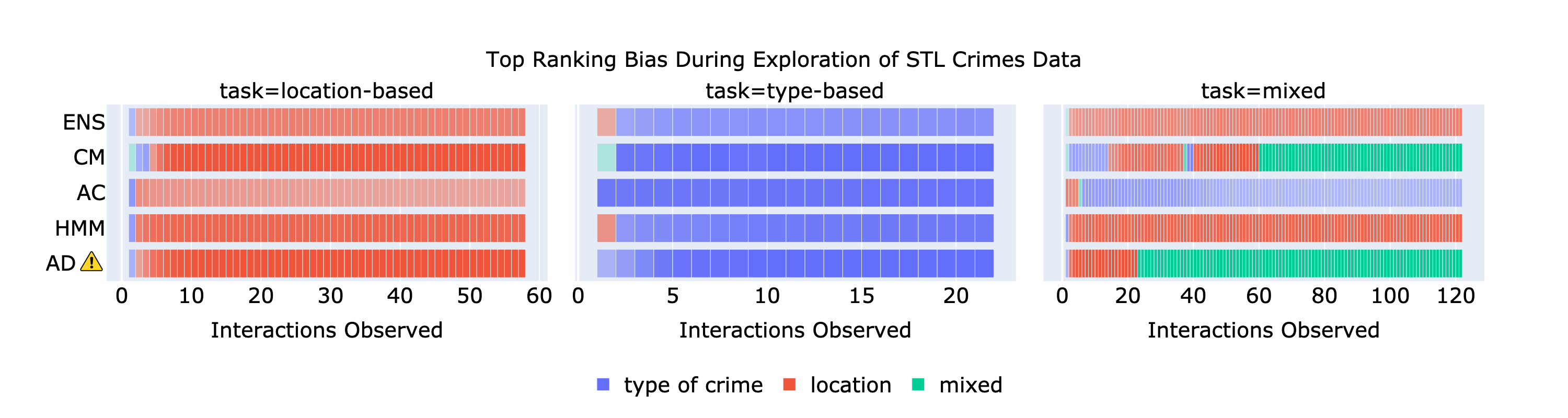}
     \end{subfigure}
     
     \caption{Each block represents a certain time $t$ within the observed interactions. The color represents the top-ranking attribute and the transparency level represents the technique's confidence. (\includegraphics[width=.015
    \textwidth]{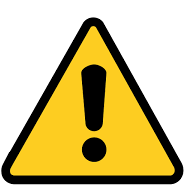} indicates that an assumption of the Chi-Square test was not met.)}
    \label{fig:bias}
    \vspace{-1em}
\end{figure*}

\section{Results}

\subsection{Data Interaction Prediction}
\label{sec:dip_results}
Figure~\ref{fig:inference} shows the models' accuracy across different values of $\kappa \in \{1,5,10,20,50,100\}$, the size of the prediction. We analyzed each algorithms' success rate at predicting the next observed interaction, calculating the overall success rate for all available user interaction traces for each dataset as \mbox{$\sum success \div \sum predictions$}. For simplicity, the results in this section focused on $\kappa = 100$, \ie\ the algorithm picks the 100 points with the highest rankings for the next interaction.

\vspace{.5em}
\noindent
\textbf{STL Crimes:}
The top row of Figure~\ref{fig:inference} shows the three subtasks in the in \textit{STL Crimes} dataset (\textit{location-based}, \textit{type-based}, and \textit{mixed}). We can observe that most of the techniques can predict the next data interaction point with high success rate at $\kappa = 100$, which represents 5\% of the underlying data. For location-based tasks, \hmm\ and \cm\ are the best performing models, with success rates of 95\% and 94\% at $\kappa = 100$, respectively. With the type-based task, \ens\ performs the best with a success rate of 93\% at $\kappa = 100$. For the mixed-based task, \ens\ outperforms all the other models, successfully predicting the next data interaction 98\% of the time. 


\vspace{.5em}
\noindent
\textbf{Vastopolis:}
The \textit{STL Crimes} user study was highly directed, asking users to perform specific tasks that lead their exploration. On the other hand, the \textit{Vastopolis} task was less directed, involved a larger data volume, and included textual data (participants interacted with Tweet-like data). The selected models are unable to reach high success rate for data interaction prediction, and success rates were consistently less than 50\% at $\kappa = 100$ points. Nonetheless, \ens\ and \af\ perform the best with this interaction log dataset. 

\vspace{.5em}
\noindent
\textbf{Boardrooms:}
The user study sessions with the \textit{Boardrooms} dataset were highly open-ended. Due to the open nature of the task, all of the models had difficulty achieving high success rate for data interaction prediction. \hmm\ outperforms all the other modeling techniques with a success rate of 51\% at $\kappa = 100$.

\vspace{.5em}
\noindent
\textbf{Political Committee:}
With the \textit{Political Committee} interaction logs, we wanted to stress-test the modeling techniques and observe their performance on set of minimal, but \textit{intentional} interactions. However, is it important to note that the dataset set is comparably small, with 180 fictitious politicians. \cm\ and \af\ showed promising performance with success rates of 67\% and 66\% at $\kappa = 100$, respectively.

\subsection{Exploration Bias Detection}
For the evaluation of bias detection, we chose the two following interaction datasets: \textit{STL Crimes} and \textit{Political Committee}. These two datasets were chosen as both studies either aimed to encourage specific biases or elicit the users' inherent biases during data exploration. Although we include the results of \ad\ in this section, we note that its performance may be unreliable as one of the Chi-Square test assumptions was not met. The Chi-Square test assumes that: \textit{the value of the expected cell should be 5 or more in at least 80\% of the cells, and no cell should have an expected of less than one}~\cite{kim2017statistical}.

\vspace{.5em}
\noindent
\textbf{Political Committee:}
Each algorithm outputs a \textit{bias value} for the attributes in the dataset. For example, \textit{Political Committee} has nine attributes including (age, gender, and occupation, \etc) and based on the assigned values, we can rank the attributes by the techniques' bias belief. For simplicity, Figure \ref{fig:bias} and the analysis in this section considers only the attribute with the highest bias belief after each user's observation. As detailed in section~\ref{sec:bias_detection_gt} our qualitative coding revealed no significant differences between the distribution of values explored by the user and the underlying data distribution, meaning no ground-truth bias was detected for every session. Although there is no evidence of exploration bias, the top of ~\autoref{fig:bias} shows the top-ranking bias found by each technique for every user throughout their data exploration. 
The absence of a ground-truth bias is potentially depicted by the lack of agreement among the techniques for the top-ranking bias. However, we see some agreement for \textit{Lima}, as the techniques believed their exploration was biased towards the occupation attribute.

\vspace{.5em}
\noindent
\textbf{STL Crimes:}
The \textit{STL Crimes} user study visually displayed three attributes (longitude, latitude, and type of crime) and included three subtasks (\textit{location-based}, \textit{type-based}, and \textit{mixed}). Since users were tasked to specifically interact with subsets of the data, we set these attributes as ground-truth biases. To calculate the bias value of location, we multiplied the bias values for longitude and latitude. To calculate the mixed bias value, we multiplied the location bias with the bias value for type of crime. The bottom row of Figure \ref{fig:bias} shows the aggregated top-ranking bias among the users over time. For both location-based and type-based tasks, all of the modeling techniques were able to accurately determine the attribute that users were most biased towards. However, for the mixed task, we observed that \ad\ and \cm\ were the only modeling techniques that detected that the users were interested in both type and location attributes. For \ens, \ac, and \hmm, the techniques believed the users were focused on only type-based and only location-based attributes, respectively.

\section{Discussion}
This study compared eight algorithms for data interaction prediction and exploration bias detection across four user study datasets. \textit{Perhaps the most salient finding is there is no clear `winner.'} In particular, we found that no single model excelled with all datasets and the differences between the top-ranking techniques are small.  
\subsection{Insights from Data Interaction Prediction Results}
The \textit{STL Crimes} data subsets demonstrate that the models generally perform relatively well when the tasks are highly-specified and have limited exploration noise. We use \textit{STL Crimes} to represent a best-case modeling scenario. Equally important, our evaluation included datasets that were selected to test the algorithms' robustness. In particular, \textit{Vastopolis} includes unstructured text, \textit{Boardrooms} involves an open exploration task, and \textit{Political Committee} presents a limited observation window with 10-16 interactions per participant. We can observe that the models' success rates fell drastically with these less directed and open-ended user study datasets. Still, there are a few noteworthy discoveries from this analysis. For example, \af~\cite{zhou2021modeling} (developed initially for textual data exploration) achieved the highest overall success rate with the \textit{Vastopolis} dataset, the only dataset with unstructured text. In addition, \hmm~\cite{ottley2019follow}, surpassed the others on the only dataset for which the task was entirely open-ended. However, in both instances, the overall success rates were relatively low.

Given the high variability in the success rates, we introduced an \ens\ approach~\cite{rokach2010ensemble,dietterich2000ensemble} by averaging the computed probabilities, allowing each model to vote for the data points they believe would best match the users' next data interactions in light of its observations. Then, we rank the data points based on their probability mass and select the \topk\ accordingly for predictions. The \ens\ model was among the top predictors for all but one dataset. We saw that averaging the predictions resulted in \ens\ achieving the second worst overall performance with the \textit{Political Committee} dataset, which we included to represent a limited observation scenario. Although we observe an overall improved performance in data interaction prediction with \ens, especially when $\kappa$ is small, we do note that this approach can be costly to compute and may not be suitable for real-time systems.
\subsection{Insights from Exploration Bias Detection Results}
The \textit{STL Crimes} dataset also provided a convenient means for evaluating exploration bias. In particular, the three data subsets involved tasks in which the study participants either focused on a specified location, data type, or a mixture of the two, allowing us to use these as ground truths. All five tested algorithms successfully identified that participants disproportionally allocated interactions toward location and type-related attributes. However, the most interesting finding relates to the mixed task data subset. Specifically, we can observe that only \cm~\cite{monadjemi2020competing} and \ad~\cite{wall2017warning}, accurately detected that the task elicited a mixed bias. 
Furthermore, the exploration bias detection findings are notably insightful for the \cm\ and \hmm\ techniques -- the only two algorithms capable of both data interaction prediction and exploration bias detection. Although both algorithms achieved more than a 90\% success rate in predicting the next data interaction at $\kappa = 100$, the exploration bias detection results provide further insight into the models' beliefs and \textit{how} they produced their data relevance rankings. Markedly, correctly identifying the mixed nature of the task meant that \cm\ was one of the best data interaction predictors, second to only \ens\ for this data subset. In contrast, misclassifying the task as location-based resulted in an overall lower prediction performance for the \hmm\ approach. 

Although our qualitative coding failed to detect exploration bias in the committee picks from participants in the  \textit{Political Committee} dataset, we included the evaluation results to assess the algorithms' predictions in light of unbiased interaction traces. We observed that all the models presented false positives, often with high certainty, as indicated by the opacity of the blocks in Figure \ref{fig:bias}. It is noteworthy that the one underlying assumption of some of the models is that the user is always biased. For example, the \hmm\ algorithm explicitly includes bias in the algorithm's specification. Specifically, it maintains a \textit{bias probability} for each feature in the dataset, allowing the algorithm to capture the likelihood of bias toward one or more data features. \cm\ is the only algorithm that directly encodes the potential for no bias. The algorithm tracks bias by building a model for all possible feature subsets, including the null set, representing no potential exploration bias. However, the null set is only one of the numerous model choices considered in \cm, making it an unlikely pick. 

Altogether, this underlying assumption that the user's interaction is always biased has both positives and negatives for real-world applications. On the one hand, we believe that identifying the features that the user disproportionately explores drives the algorithms' success in data interaction prediction. Also, it is potentially a reasonable assumption for many exploration scenarios. On the other hand, the false positives can erode the user's trust in the algorithms, especially in proposed applications related to helping the user recognize their own exploration bias or recommender systems~\cite{gotz2016adaptive, wall2017warning, wall2022left}. That said, a potential solution exists within two models, \ac~\cite{gotz2016adaptive} and \ad~\cite{wall2017warning}, which leverage statistical methods for hypothesis testing in their bias detection. If used as intended, a failure to reject the null hypothesis provides evidence of no bias. However, practitioners and researchers should use the appropriate tests for the given data and verify that the data meet their assumptions.

\subsection{Challenges and Limitations}
There were several challenges we faced when preparing to perform a unified comparison of these models. For example, implementations of the techniques were not readily available. We invested time gaining an understanding of all the techniques and standardizing the implementations to handle various datasets. We note that we are not able to completely replicate the performance of \cm\ and \hmm\ on data interaction prediction with the \textit{STL Crimes} dataset that we see in Monadjemi et al~\cite{monadjemi2020competing}. We believe that the differences in performance are likely due to our hyperparameters selections.
When deploying algorithms in real-world settings, it is important to gather data and tune hyperparameters. However, we excluded this step in our work as the datasets were small, presenting a high risk for overfitting. Instead, we used our domain knowledge to make conscientious decisions about each model's hyperparameters, which are all available and adjustable in our codebase. 

Although we selected a diverse collection of datasets, we acknowledge that they do not represent many visual analytics scenarios. Therefore, we take care in reporting our observations and hesitate to make broad statements. Further, we were limited by the small number of interaction datasets that are publicly available, especially those with ground truths. This highlights important considerations for balancing ecological validity where the ground truth is not known or may not exist, and having some measure of success for evaluation. Determining ground truth when evaluating bias in open-ended tasks is not trivial. A call for more guided user studies with built-in biases may be needed so that researchers can further evaluate bias detection algorithms.


\section{Future Directions}
Despite our limitations, our work is a step forward in evaluating user modeling techniques proposed in the visual analytics community and opens possibilities for future work in this area. For example, our evaluations only involved datasets with a single interaction type, even though some of the modeling techniques are able to or have the potential to handle multiple types of interactions~\cite{zhou2021modeling, ottley2019follow, monadjemi2020competing}. Although our implementations did not include this feature, future work could extend this benchmark study with the added ability to learn from multiple types of interactions with data points. Further, we primarily focused on quantitative measures, but there are other factors beyond accuracy that might affect the appropriateness of a technique for a given user modeling scenario. Factors such as speed, trustworthiness, and flexibility can be important considerations to explore in future evaluations of user modeling techniques.

\section{Conclusion}
In this paper, we present a computational benchmark study that provides a unified comparison of user modeling techniques for data interaction prediction and exploration bias detection. We implemented standardized versions of seven previously proposed user modeling techniques that learns from a user's low-level interactions in real-time. Additionally, we developed an ensemble approach by averaging the models' predictions. We then evaluated the performance of all these techniques across four different interaction logs for data interaction prediction/data ranking and exploration bias detection. We found that there is no clear `winner' among the modeling techniques, but our analysis highlights the tasks and datasets that elicited the best performance for each technique. Finally, we discuss the open challenges for user modeling and evaluations. This work is a step towards gaining an understanding of user modeling techniques within the visual analytics community and we hope that it encourages further advances in analyzing user interactions and visualization provenance.

\acknowledgments{
This work is supported in part by the National Science Foundation under Grant No. OAC-2118201 and IIS-2142977.}
\clearpage
\bibliographystyle{abbrv-doi}

\bibliography{references}
\end{document}